 \documentstyle[twoside,fleqn,espcrc2,epsf]{article}

\newcommand{\AmS}{{\protect\the\textfont2
  A\kern-.1667em\lower.5ex\hbox{M}\kern-.125emS}}

\newcommand{\be}{\begin{equation}}
\newcommand{\ee}{\end{equation}}
\newcommand{\beq}{\begin{eqnarray}}
\newcommand{\eeq}{\end{eqnarray}}

\title{The Matter Density Distribution for Mesons and Baryons
\thanks{Presented by C.~Alexandrou}}

\author{C. Alexandrou\address[Cyprus]{Department of Physics, University of Cyprus,
CY-1678 Nicosia, Cyprus},
Ph.\ de Forcrand\address{ETH-Z\"urich, CH-8093 Z\"urich and CERN Theory Division, CH-1211 Geneva 23, Switzerland} 
and A. Tsapalis\thanks{Supported by the Levendis Foundation}\addressmark[Cyprus] }
       
\begin{document}

\begin{abstract}
The matter density distribution inside a hadron is evaluated using gauge-invariant 
correlation functions within the quenched and the
unquenched theory. 
Comparison with the charge density distribution suggests that 
hadron deformation is a consequence of the relativistic motion of the quarks.

\vspace{1pc}
\end{abstract}

\maketitle

\vspace*{-1.8cm}

\section{Introduction}

 The charge and matter density distributions
encode detailed information on hadron structure.
 In the non-relativistic limit both charge and matter density
distributions reduce to the square of the wave function. Therefore
differences  between them give a measure of relativistic effects.
 Hadron deformation can be searched via the matter 
density distribution  as well as via the charge density distribution.
There are strong experimental indications from measurements of the 
electromagnetic transition form factors in $\gamma^* N$ to $\Delta$~\cite{Bates}
that the nucleon is deformed.
It is thus interesting to address the issue of hadron deformation within
lattice QCD and understand its physical origin. Comparison
of matter and charge density distributions can help discriminate
among possible mechanisms.
 A number of phenomenological studies use  the bag-model 
to describe hadron deformation. For this reason we compare the
charge and  matter distributions to the
bag-model results.

\vspace{-0.3cm}

\section{Observables}

We evaluate  equal time two-current correlators 

\vspace*{-0.3cm}
\be
C_\Gamma({\bf r},t) = \int\> d^3r'\>
\langle h|j_\Gamma^{u}({\bf r}'+{\bf r},t)j_\Gamma^{d}({\bf r}',t)|h\rangle
\label{correlator}
\ee

\vspace*{-0.3cm}
\noindent
with the current operator  given by the normal order product
$j_\Gamma^u({\bf r},t)=:\bar{u}({\bf r},t)\Gamma u({\bf r},t):$ and
$\Gamma=\gamma_0, 1$
for the charge and matter correlators respectively.
We note that no gauge ambiguity arises in
this determination of hadron wave functions 
unlike Bethe-Salpeter amplitudes.
In the case of  baryons two relative distances are
involved and three current insertions
are required. 
However we may consider integrating over one relative 
distance
to obtain the one-particle density distribution that involves two current
insertions  and it is thus evaluated via Eq.~\ref{correlator}.

\vspace*{-0.3cm}

\section{Results}

All the results presented here have been obtained on  lattices of size
$16^3 \times 32$. We analyze, for the quenched case, 220 NERSC 
configurations  at $\beta=6.0$ 
 for pion to rho mass ratio  $m_\pi/m_\rho=0.88, \>0.84,\>0.78$ and 0.70 and, for the unquenched,
150 SESAM configurations~\cite{SESAM} for $\kappa=0.156$  and 200 
for $\kappa=0.157$ with $m_\pi/m_\rho=0.83$ and $0.76$ respectively.
The physical volume of the quenched and unquenched lattices
is approximately the same.

In Eq.~\ref{correlator} the current couples to the quark at fixed time separation $t$ from the source,
which must be  large enough to 
sufficiently isolate the hadron ground state. 
Using local and Wuppertal smeared sources
we first checked that we obtain  the same value for the effective mass 
at $t=8a$. This value  agrees with the one extracted 
using data with
 Dirichlet boundary conditions 
utilizing the whole time extent of our lattice.
 In contrast,  wall sources  
even at maximal separation $t=8a$  
still show a
sizable excited states contamination.  

\begin{figure}[h]
\vspace*{-0.8cm}
\mbox{\epsfxsize=7truecm\epsfysize=5.5truecm\epsfbox{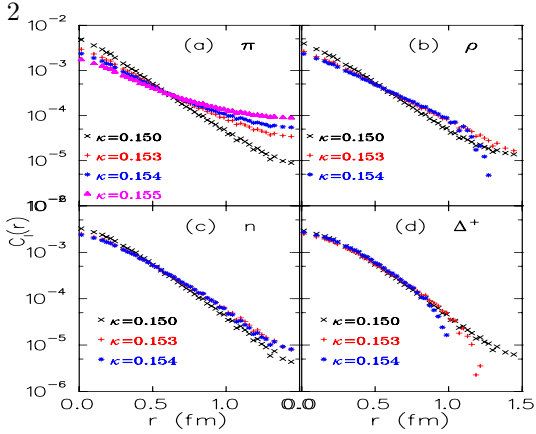}}
\vspace*{-1cm}
\caption{$C_I(r)$ for different quark masses. The errors bars are not shown for clarity.}
\vspace*{-0.7cm}
\label{matter}
\end{figure}

\noindent
In addition, we checked, by comparing data  at different $t$, 
that higher momenta 
are sufficiently suppressed in  the current-current correlators
since summing over the spatial volume of
the  source is not possible.
Also,  results obtained using Wuppertal smearing which suppresses
 high momenta and
local sources which allow them, are in agreement at $t=8a$.
This means that  contamination  from high momentum states
is negligible  at $t=8a$ allowing us to analyze standard full QCD
configurations that
employ anti - periodic b.c.
Since, for our
parameters,   local sources produce the same results for $C_\Gamma(r)$ 
as smeared ones and carry no gauge noise they
are preferable for this study.

The quenched matter density distribution shown in Fig.~\ref{matter}
broadens for the pion as the quarks become lighter.
For the rho, the nucleon and the $\Delta^+$ 
essentially no variation is seen
over  the range of naive quark masses $\sim  300-100$~MeV.

\begin{figure}[h]
\vspace*{-0.8cm}
\mbox{\epsfxsize=7truecm\epsfysize=4.5truecm\epsfbox{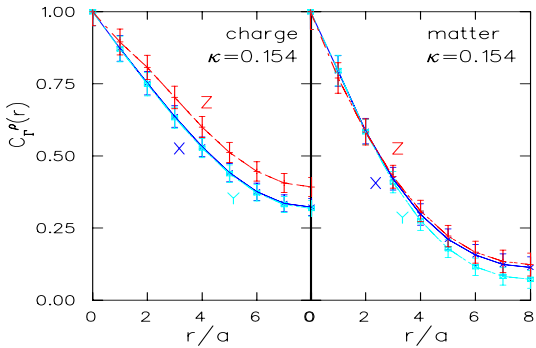}}
\vspace*{-1.cm}
\caption{$C_\Gamma(0,0,z)$ (upper Z-curve) and
$C_\Gamma(x,0,0)$ and $C_\Gamma(0,y,0)$ (lower $X,Y$-curves).}
\label{rho-asymmetry}
\vspace*{-0.7cm}
\end{figure}

\begin{figure}[h]
\vspace*{-0.8cm}
\mbox{\epsfxsize=7truecm\epsfysize=6truecm\epsfbox{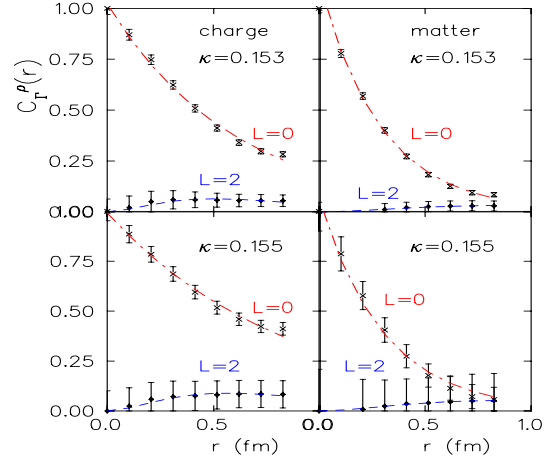}}
\vspace*{-1cm}
\caption{
 Decomposition into 
 $L=0$ and $L=2$ contributions. 
}
\vspace*{-0.7cm}
\label{deformation}
\end{figure}

In a previous study of the charge density distribution~\cite{AFT,panic02}
we obtained a non-zero quadrupole moment for the rho
or a $z-x$ asymmetry where the $z$-axis is taken along
the spin of the rho as shown in Fig.~\ref{rho-asymmetry}. 
 An angular decomposition 
 of the wave function, as shown in Fig.~\ref{deformation}, corroborates
 a non-zero charge deformation by
 producing a non-zero $L=2$ component.
No such  deformation is 
seen  for the matter density distribution in Figs.~\ref{rho-asymmetry},~\ref{deformation} .
 Since the matter and charge operators
have the same  non-relativistic limit,
this strongly suggests  that hadron
charge deformation is a relativistic effect.
This result has
 strong implications for the validity of various phenomenological models used
in the study of nucleon deformation and  
 $\gamma^{\star} N\rightarrow\Delta$
form factors.

\begin{figure}[h]
\vspace*{-0.8cm}
\mbox{\epsfxsize=7truecm\epsfysize=4.truecm\epsfbox{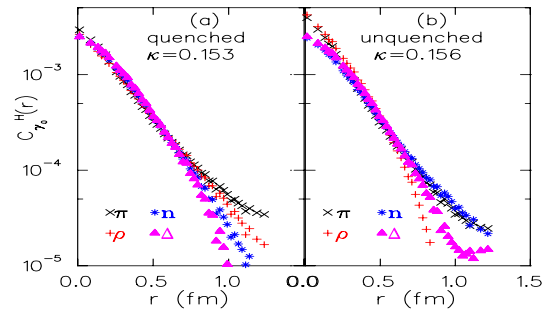}}
\vspace*{-1cm}
\caption{(a) Quenched  and (b) unquenched  matter density  distributions.}
\vspace*{-0.75cm}
\label{full}
\end{figure}

Fig.~\ref{full} compares quenched and unquenched results for $C_I(r)$ at $m_\pi/m_\rho \approx 0.83$.
Unquenching increases $C_I(r)$ at short
distances in the case of the pion and the rho whereas for the baryons
 no significant changes are seen.
Both quenched and unquenched matter distributions
show a faster fall off 
than the charge distribution.
Furthermore,  whereas unquenching
tends to lead to a small increase in the  rho charge asymmetry, 
it has no effect on the matter density distribution.
This again suggests a relativistic origin for the rho charge deformation. 

\vspace*{-0.3cm}

\section{Comparison of lattice and  bag model results}

\vspace*{-0.3cm}

 We consider only the lowest mode
of the free Dirac field in a spherical bag of radius $R$
that is chosen so as to minimize the mass
 of the hadron
under consideration.  For the
comparison presented here,  we use 
the lattice values for the masses of the rho, the nucleon and the
$\Delta$ to fix the bag  parameters,  
$B, Z_0$ and   $\alpha_{bag}$,
 using as an input
 the naive quark mass. 
Expanding the quark fields in terms of bag eigenmodes 
we obtain for 
 the charge and matter  density distributions 

\vspace*{-0.2cm}
\be
\left \langle
\hat{j}_{\gamma_0}^u({\bf r})\>  \hat{j}_{\gamma_0}^d({\bf r}')  \right \rangle
_{M \atop B} = \mp C 
\> \begin{array}{c}\left(f(r)^2+g(r)^2 \right) \\
\left( f(r')^2+g(r')^2 \right)
\end{array}
\ee

\vspace*{-0.5cm}

\be
\left \langle
  \hat{j}_{I}^u({\bf r})\>  \hat{j}_{I}^d({\bf r}')    \\
\right \rangle_{M \atop B}
=  C' 
 \begin{array}{c} \left( f(r)^2-g(r)^2 \right)\\
 \left( f(r')^2-g(r')^2 \right)
\end{array}
\ee
with  $C,\> C'$ 
 constants.
$M$ denotes mesons and 
$B$ baryons.
These expressions show explicitly that the difference between charge and matter
density distributions is due to the opposite sign of the lower  components
of the Dirac spinor i.e. a relativistic effect.

\begin{figure}[h]
\vspace*{-0.7cm}
\mbox{\epsfxsize=7truecm\epsfysize=6.4truecm\epsfbox{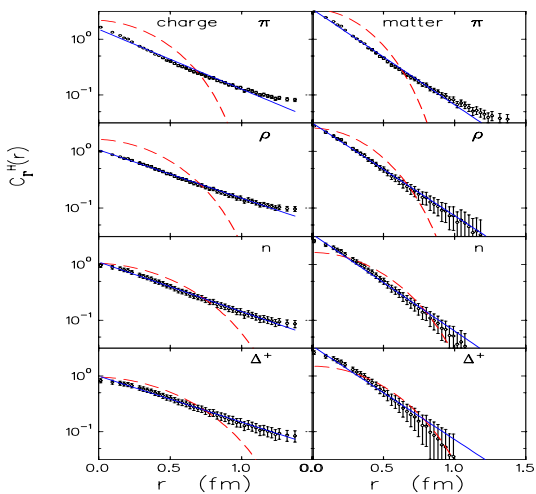}}
\vspace*{-1cm}
\caption{Quenched charge and matter distributions 
 at $\kappa=0.153$ and bag model results (dashed line).
The solid line is a fit to the form $\exp(-m r)$.}
\vspace*{-0.8cm}
\label{bag model}
\end{figure}

From Fig.~\ref{bag model} we see that the bag model 
fails  to reproduce the correct radial dependence of the
lattice results for the charge
and matter density distributions 
 and that the Ansatz $\exp(-mr)$
 provides the best description of the data. 
A measure
of the width of the distributions is provided by the  root mean square (rms) radius.
The ratio of the charge to the matter rms radius
in the bag-model at $\kappa=0.153$ and $0.154$ is: $1.16$ and $1.17$
for the rho and $1.16$ 
for the nucleon.
The corresponding ratios for the lattice data 
are $1.15(3)$ and $1.16(8)$ for the rho
and $1.18(4)$ and $1.20(8)$ for the nucleon.
Thus, despite the failure of the bag model in describing the 
individual radial shape
of the distributions
it produces reasonable results for the relative  widths of
 the charge to the matter distribution.
 Both lattice and bag model
consistently predict
a broader charge than matter distribution,
with a very weak mass dependence.

\vspace*{-0.2cm}

\section{Conclusions}

\vspace*{-0.2cm}

 For the lattice parameters
used in  this study,
local sources are suitable interpolating fields
  for the evaluation of density
distributions,
since they have less gauge
noise than smeared ones and
the temporal extent of our lattice is large enough  
 to obtain the ground state.
Quenched and unquenched
results for both the charge and matter 
density distributions show insignificant differences.
 The
charge density distribution is, in all cases,
broader than the  matter density. For baryons, the lattice indicates a charge  radius about
20\% larger than the  matter radius.
This effect is well reproduced by the bag model. 
 The deformation seen in the rho charge distribution 
is absent in the matter
distribution, both in the quenched and the unquenched theory. This
observation suggests a relativistic origin for the deformation.

\end{document}